\documentclass[a4paper,11pt]{article}
\pdfoutput=1 
\usepackage{jcappub} 
\usepackage[T1]{fontenc} 
\usepackage{amsmath}
\usepackage{xcolor}
\usepackage{color}
\usepackage{soul,xcolor}
\usepackage[normalem]{ulem} 
\setstcolor{red}
\title{\boldmath Cosmological Dynamics of Relativistic MOND}

\author[a]{Tahere Kashfi}
\author[a,b,1]{, Mahmood Roshan\note{Corresponding author}}

\affiliation[a]{Department of Physics, Faculty of Science, Ferdowsi University of Mashhad, P.O. Box 1436, Mashhad, Iran}
\affiliation[b]{School of Astronomy, Institute for Research in Fundamental Sciences (IPM), P. O. Box 19395-5531, Tehran, Iran}

\emailAdd{stkashfi@mail.um.ac.ir}
\emailAdd{mroshan@um.ac.ir}

\abstract{In this paper, we investigate the recently proposed relativistic theory for MOND using the phase space analysis. Unlike its precedent theory, namely TeVeS, this theory is claimed to agree with the observed Cosmic Microwave Background and matter power spectra. We convert the background cosmological equations of the theory to a set of first-order autonomous equations. Then we explore the corresponding fixed points and their physical meaning. This method is powerful in the sense that the cosmological behavior of the model is diagnosed independently of the magnitude of the free parameters of the theory. We show that the theory has a viable sequence of cosmological epochs. Although this theory mimics the standard cosmological model when specific conditions are used, it provides a richer structure as far as the background cosmology is concerned. This implies that further investigations are required to see if this theory contributes to addressing the current cosmological tensions.}

\begin{document} 	
\maketitle
\flushbottom
\section{Introduction}\label{sec:intro}

The dark matter particles have not yet been detected. Therefore, alternative theories of gravity are still a possible approach to address the missing mass problem. Although modified gravity theories can explain the flat rotation curves of the spiral galaxies and some other local properties of different types of galaxies, the formation of the cosmic structures has been a serious challenge for them. In other words, although most of the modified gravity theories that deny the existence of the cold dark matter particles can be compatible with the local/galactic scale observations, they fail in explaining the cosmic observations like Cosmic Microwave Background (CMB) and matter power spectra observations. 

Modified Newtonian dynamics (MOND) is one of the phenomenological theories that are successful in explaining many properties of the galaxies \cite{Milgrom:1983pn}. There is a universal acceleration scale $a_0\simeq 1.2\times 10^{-10}\,\text{m}\text{s}^{-2}$ in this theory beyond which the MOND effects appear, see \cite{Famaey:2011kh} and \cite{Banik:2021woo} for reviews on MOND. There have been several attempts to construct a relativist theory for MOND. Relativistic Aquadratic Lagrangian (RAQUAL) proposed in \cite{Bekenstein:1984tv} for the first time. The scalar field in RAQUAL plays the role of an auxiliary potential, and its gradient then has the dimensions of acceleration and can be used to enforce the acceleration-based modification of MOND. However, gravitational lensing is insensitive to such conformal rescalings of the metric (apart from the contribution of the stress-energy of the scalar field itself), and the non-Newtonian effects of the theory are always very different on dynamics (which is MONDian) and lensing (which is not). A solution to this problem was therefore proposed in \cite{Sanders:1996wk}, inspired by 'stratified' theories of gravity. More specifically, in addition to the scalar field of RAQUAL, one may use a non-dynamical time-like vector field with unit-norm in order to have a disformal rescaling of the metric. Endowing this vector field with covariant dynamics of their own has then been the next logical step in developing relativistic MOND theories, which was actually first achieved in \cite{Bekenstein:2004ne}. Tensor-Vector-Scalar (TeVeS) has been a proof of concept that it is possible to construct a fully covariant theory reproducing MOND in the weak-field limit for both dynamics and lensing. It was, however, quickly shown that this theory had a Hamiltonian density unbounded by below and that, even at the classical level, spherically-symmetric solutions were unstable. A generalized version of TeVeS, not suffering from similar theoretical problems, was later proposed in \cite{Skordis:2008pq}, but the whole framework was ruled out by actual observations, in particular by the observations of GW170817 followed by an electromagnetic counterpart, demonstrating the equality of the speed of light and gravity. Also, the theory fails in explaining the CMB observations \cite{Dodelson:2006zt}. This was then cured in a new version of a TeVeS-like theory by \cite{Skordis:2019fxt}. However, it should be noted that if sterile neutrinos with the mass of 11 $\text{eV}/c^2$ are added with the same relic abundance as the  cold dark matter in the standard $\Lambda$ Cold Dark Matter ($\Lambda$CDM) model, then the CMB can be fit within TeVeS \cite{Angus:2008qz}.

Recently, a new relativistic theory for MOND (RMOND) has been proposed in \cite{Skordis:2020eui}. This theory shares some common features with TeVeS. The theory is claimed to be consistent with CMB and matter power spectra observations. Although there is no dark matter particle postulated in this theory, the corrections that are induced to the standard Friedman equations via the existence of the extra vector and scalar fields may mimic the dark matter behavior. 

In this paper, we investigate the background cosmology in the context of RMOND. We use the dynamical system approach in order to diagnose the general cosmological behavior of the model. The outline of the paper is as follows: in Sec. \ref{FE} we review the field equations. In Sec. \ref{FRE}, we derive the modified version of the Friedman equations. In Sec. \ref{DS}, we apply the dynamical system approach to RMOND by defining a suitable set of phase space variables. Then we discuss the meaning of the fixed points and the cosmological viability of RMOND. The conclusions are drawn in Sec. \ref{CO}.

\section{The field equations of RMOND}\label{FE}

The generic action of the theory is given by $S=S_G+S_M$, where $S_M$ is the action of the ordinary matter and $S_G$ is given by \cite{Skordis:2020eui}
\begin{equation}
	S_G=\int d^4 x \frac{\sqrt{-g}}{16\pi G} \Big[R-\frac{K_B}{2}F_{\mu\nu}F^{\mu\nu}+2(2-K_B)J^{\mu}\nabla_{\mu}\phi-\mathbb{F}(\mathcal{Y},Q)-\hat{\lambda}(A^{\mu}A_{\mu}+1)\Big]
\end{equation}
where $\hat{\lambda}$ is the Lagrange multiplier implying the constraint that $A^{\mu}A_{\mu}=-1$. Only the metric tensor, not the other fields, appears in $S_M$. Therefore the standard conservation equation for the energy-momentum tensor $T_{\mu\nu}$ of the normal matter is satisfied. The scalars $Q$ and $\mathcal{Y}$, and the vector current $J^{\mu}$ are defined as
\begin{equation}
	Q=A^{\mu}\nabla_{\mu}\phi,\,\,\,\,\,\,\,\,\,\, \mathcal{Y}=(g^{\mu\nu}+A^{\mu}A^{\nu})\nabla_{\mu}\phi \nabla_{\nu}\phi,\,\,\,\,\,\,\,\,\,\, J^{\mu}=A^{\alpha}\nabla_{\alpha}A^{\mu}
\end{equation}
In the original notation used in \cite{Skordis:2020eui}, the function $\mathbb{F}$ takes the following form
\begin{equation}
	\mathbb{F}(\mathcal{Y},Q)=2\Lambda+(2-K_B)\mathcal{Y}+\mathcal{F}(\mathcal{Y},Q)
\end{equation}
where $\Lambda$ is the cosmological constant responsible for the cosmic speed up. As we will see, the new fields of RMOND cannot play the role of the dark energy. So it is necessary to keep $\Lambda$. Now let us assume a general form for the $\mathcal{F}(\mathcal{Y},Q)$ as
\begin{equation}
	\mathcal{F}(\mathcal{Y},Q)=-2\mathcal{K}(Q)+\sum_{n=2}^{\infty}b_n \mathcal{Y}^n
\end{equation}
where $b_n$ are constant coefficients. On the other hand, let us define the functions $\mathbb{F}_{Q}$ and $\mathbb{F}_{\mathcal{Y}}$ as follows \begin{equation}
	\mathbb{F}_{Q}=\frac{\partial \mathbb{F}}{\partial Q}=-2\frac{d\mathcal{K}}{d Q}=-2 \mathcal{K}_Q,\,\,\,\,\,\,\,\,\,\,\, \mathbb{F}_{\mathcal{Y}}=\frac{\partial \mathbb{F}}{\partial \mathcal{Y}}
\end{equation}
Now by varying the action $S$ with respect to the metric tensor $g_{\mu\nu}$, we find the following field equation
\begin{equation}\label{E1}
	G_{\mu\nu}+\mathcal{H}_{\mu\nu}=\kappa T_{\mu\nu}
\end{equation}
where $G_{\mu\nu}$ is the Einstein tensor and $\kappa=8\pi G$. We use the units in which the speed of light is $c=1$. On the other hand, the tensor $\mathcal{H}_{\mu\nu}$ includes all the corrections of this theory to General Relativity (GR) and is given by
\begin{equation}
	\begin{split}
		\mathcal{H}_{\mu\nu}=&-K_B(F_{\mu\sigma}F_{\nu}^{~\sigma}-\frac{1}{4}g_{\mu\nu}F^{\alpha\beta}F_{\alpha\beta})-\hat{\lambda} A_{\mu}A_{\nu} -(2-K_B)g_{\mu\nu}J^{\alpha}\nabla_{\alpha}\phi\\&
		+2(2-K_B)(A^{\sigma} \nabla_{(\mu}\phi \nabla_{\sigma}A_{\nu)}-\frac{1}{2}A_{\mu}A_{\nu}\Box\phi+\nabla_{\sigma}\phi A_{(\mu}F_{\nu)}^{~\sigma})\\&
		+\frac{1}{2} g_{\mu\nu}\mathbb{F}-\mathbb{F}_{\mathcal{Y}} \nabla_{\mu}\phi \nabla_{\nu}\phi-(2 Q\mathbb{F}_{\mathcal{Y}}+\mathbb{F}_Q)A_{(\mu}\nabla_{\nu)}\phi
	\end{split}
\end{equation}

Variation of the action with respect to the vector field yields 
\begin{equation}\label{E2}
	K_B\nabla_{\mu}F^{\mu\nu}+(2-K_B)(\nabla_{\mu}\phi\,\nabla^{\nu}A^{\mu}-\nabla_{\mu}(A^{\mu}\nabla^{\nu}\phi))-\hat{\lambda} A^{\nu}-\frac{\nabla^{\nu}\phi}{2}(2 Q\mathbb{F}_{\mathcal{Y}}+\mathbb{F}_Q)=0
\end{equation}
and the field equation of the scalar field $\phi$ is
\begin{equation}\label{E3}
	\nabla_{\mu}(\mathbb{F}_Q A^{\mu})-2(2-K_B)\nabla_{\mu}J^{\mu}+2\nabla_{\mu}(Q\mathbb{F}_{\mathcal{Y}}A^{\mu})+2\nabla_{\mu}(\mathbb{F}_{\mathcal{Y}} \nabla^{\mu}\phi)=0
\end{equation}
Although with RMOND, we deal with a classical theory of gravity, it is natural to ask: what is the difference between adding new fields to the action of the gravitational theory and postulating new dark matter particles? If there is no difference, why should we call RMOND a modified theory of gravity? Let us explain why there are some basic differences. In galactic systems, the RMOND fields do not behave by their energy (mass) density to strengthen gravity, as do dark matter particles. Instead, these fields modify the gravitational interactions between bodies. Another reason that these fields do not act like dark matter particles is that they are produced by the baryons and stick with them. This means they cannot be removed from a self-gravitating system leaving a bare baryonic system. However, this is quite possible for dark matter particles via mechanisms like tidal stripping.

\section{The modified Friedmann equations in RMOND}\label{FRE}
In this section, we find the modified Friedmann equations in RMOND. To do so, we start with the flat Friedmann–Robertson–Walker (FRW) metric
\begin{equation}
	ds^2=-dt^2+a(t)^2(dr^2+r^2 d\Omega^2)
\end{equation}
where $a(t)$ is the cosmic scale factor. In the FRW space-time, all the fields depend only on the cosmic time, and the above-mentioned complicated field equations can be substantially simplified. More specifically, it is straightforward to show that $A_{\mu}=(-1,0,0,0)$, $F_{\mu\nu}=0$, $ \mathcal{Y}=0$, $J^{\mu}=0$, $Q=\dot{\phi}$, and 
\begin{equation}
	\begin{split}
		&\mathbb{F}=2\Lambda-2\mathcal{K},\\&\mathbb{F}_{\mathcal{Y}}=(2-K_B),\\& \mathbb{F}_Q=-2 \mathcal{K}_Q
	\end{split}
\end{equation}
It is easy to show that the field equation \eqref{E3} reduces to $\nabla_{\mu}(\mathbb{F}_Q A^{\mu})=0$.  This equation in the FRW space-time takes the following form
\begin{equation}\label{E4}
	\dot{\mathcal{K}}_Q+3\, H{\mathcal{K}}_Q=0
\end{equation}
where $H(t)=\dot{a}/a$ is the Hubble function. This equation shows that $\mathcal{K}_Q$ drops as $1/a^3$. This is one of the main features of the theory and implies that some contributions of the extra fields in this theory behave like cold dark matter. This is similar to what happens in mimetic gravity \cite{Chamseddine:2013kea}. On the other hand, we use the field equation of $A^{\mu}$, namely \eqref{E2}, to fix the Lagrange multiplier $\hat{\lambda}$. The result is
\begin{equation}
	\hat{\lambda}=(2-K_B)(\dot{Q}+3 H Q+Q^2)-Q \mathcal{K}_Q
\end{equation}
By substituting this equation into \eqref{E1}, and assuming that the cosmic fluid is described by an ideal fluid with the baryonic density $\rho$ and pressure $p$, we find the generalized Friedmann equations in RMOND:
\begin{equation}\label{E5}
	H^2=\frac{\kappa}{3}\rho+\frac{\Lambda}{3}+\frac{1}{3}(Q\mathcal{K}_Q-\mathcal{K})
\end{equation}
\begin{equation}\label{E6}
	\dot{H}+H^2=-\frac{\kappa}{6}(\rho+3 p)+\frac{\Lambda}{3}-\frac{1}{6}(Q\mathcal{K}_Q+2\mathcal{K})
\end{equation}
 These equations combined with \eqref{E4} make the governing equations for the dynamics of the background cosmology. It is necessary to mention that equations \eqref{E5} and \eqref{E6} take the standard form as in $\Lambda$CDM by defining the following effective density and pressure
\begin{equation}\label{E7}
	\hat{\rho}=\frac{1}{\kappa}(Q\mathcal{K}_Q-\mathcal{K}), \,\,\,\,\,\,\,\,\,\,\,\,\, \hat{p}=\frac{\mathcal{K}}{\kappa}
\end{equation}
Although $\mathcal{K}_Q$ drops as $1/a^3$, it is clear that $\hat{\rho}$ does not behave like the cold dark matter component in $\Lambda$CDM. Therefore, it is not trivial if the cosmic history in this theory possesses a true sequence of cosmic epochs. This is why the dynamical system analysis of the theory may provide valuable information about the validity of the theory.

It is crucial to know the function $\mathcal{K}$. The following four functions in connection with CMB observations have been proposed in \cite{Skordis:2020eui}:
\begin{equation}\label{E8}
	\mathcal{K}_1(Q)={\kappa_2}(Q-Q_0)^2
\end{equation}
\begin{equation}\label{E9}
	\mathcal{K}_2(Q)=\frac{\kappa_2}{4 Q_0^2}(Q^2-Q_0^2)^2
\end{equation}
\begin{equation}\label{E10}
	\mathcal{K}_3(Q)=2\kappa_2\mathcal{Z}_0^2(\cosh\mathcal{Z}-1)
\end{equation}
\begin{equation}\label{E11}
	\mathcal{K}_4(Q)=2\kappa_2\mathcal{Z}_0^2(e^{\mathcal{Z}^2}-1)
\end{equation}
where $Q_0$, $\kappa_2$ and $\mathcal{Z}_0$ are free parameters that should be fixed by cosmic observations, and $\mathcal{Z}$ is defined as $\mathcal{Z}=(Q-Q_0)/\mathcal{Z}_0$. As already mentioned, irrespective of the functional form of $\mathcal{K}(Q)$, its derivative with respect to $Q$ drops as $\mathcal{K}_Q=I_0/a^3$. Where $I_0$ is a constant of integration. This means that it is possible to find $Q$ in terms of $a$, and consequently $\hat{\rho}$ as a function of the scale factor. It is easy to verify that the corresponding effective densities are
\begin{equation}\label{E12-1}
	\begin{split}
		&\kappa\hat{\rho}_1(a)=\frac{I_0 Q_0}{a^3}+\frac{I_0^2}{4\kappa_2 a^6}\\&
		\kappa\hat{\rho}_2(a)\simeq \frac{I_0 Q_0}{a^3}+\frac{I_0^2}{4\kappa_2 a^6}-\frac{I_0^3}{8\kappa_2^2 Q_0 a^9}+\frac{I_0^4}{8\kappa_2^3 Q_0^2 a^{12}}+\mathcal{O}(I_0^5)\\&
		\kappa\hat{\rho}_3(a)\simeq \frac{I_0 Q_0}{a^3}+\frac{I_0^2}{4\kappa_2 a^6}-\frac{I_0^4}{192\kappa_2^3 \mathcal{Z}_0^2 a^{12}}+\frac{I_0^6}{2560\kappa_2^5  \mathcal{Z}_0^4 a^{18}}+\mathcal{O}(I_0^8)\\&
		\kappa\hat{\rho}_4(a)\simeq \frac{I_0 Q_0}{a^3}+\frac{I_0^2}{8\kappa_2 a^6}-\frac{I_0^4}{256\kappa_2^3 \mathcal{Z}_0^2 a^{12}}+\frac{5I_0^6}{12288\kappa_2^5  \mathcal{Z}_0^4 a^{18}}+\mathcal{O}(I_0^8)
	\end{split}
\end{equation}
In all cases, the first term on the right-hand side is $\frac{I_0 Q_0}{a^3}=Q_0 \mathcal{K}_Q$ and mimics the  cold dark matter behavior. Now the question is: what is the role of the other terms proportional to $a^{-n}$ with $n>3$? The effective pressure $\hat{p}$ is also nonzero and can be similarly expanded over the scale factor $a$. This directly means that the equation of state parameter $\hat{\omega}=\hat{p}/\hat{\rho}$ varies with time and may cause significant deviations from standard cosmology. Although these extra terms will decay in the late time universe, their impact on the early or intermediate universe needs to be carefully checked. In the next section, we explore this concern.

\section{RMOND as a dynamical system}\label{DS}
The background cosmology in RMOND is described by three equations \eqref{E4}, \eqref{E5} and \eqref{E6} alongside the suitable equation of state for the normal matter distribution. In this section, by choosing appropriate phase space variables, we convert the governing equations to a system of autonomous first-order dynamical system differential equations. For a comprehensive review of the dynamical system approach in the cosmological models, we refer the reader to \cite{Bahamonde:2017ize}. 

Our analysis here is independent of the form of the main function $\mathcal{K}(Q)$. However, the decomposition based on the different functions of $\mathcal{K}(Q)$ presented at the end of the previous section, i.e., equation \eqref{E12-1} is helpful to define the phase space variables. The density $\rho$ can be written as the combination of the non-relativistic $\rho_m$ and relativistic $\rho_r$ components. We combine $\rho_m$ with $\rho_d=Q_0 \mathcal{K}_Q/\kappa$ in the dimensionless variable $\Omega_m(t)$ as
\begin{equation}
	\Omega_m=\frac{\kappa}{3 H^2}(\rho_m+\rho_d)
\end{equation}
notice that $\rho_d$ is the effective dark matter introduced by RMOND. Therefore the total non-relativistic "matter" contribution to the energy-mass budget of the cosmos is identified by $\Omega_m$. The other variables are defined as
\begin{equation}
	\Omega_r=\frac{\kappa\rho_r}{3 H^2}, \,\,\,\,\, \Omega_{\Lambda}=\frac{\Lambda}{3 H^2}, \,\,\,\,\,\Omega_{y}=\frac{\mathcal{K}}{3H^2}, \,\,\,\,\,\Omega_x=\frac{(Q-Q_0)\mathcal{K}_Q-\mathcal{K}}{3H^2}, \,\,\,\,\,\lambda=\frac{\mathcal{K}_Q^2}{\mathcal{K}_{QQ}\mathcal{K}}
\end{equation}
where $\mathcal{K}_{QQ}=\frac{d^2\mathcal{K}}{dQ^2}$. With these variables, the equation \eqref{E5} takes the following form
\begin{equation}\label{E12}
	\Omega_m+\Omega_r+\Omega_{\Lambda}+\Omega_x=1
\end{equation}
Notice that $\Omega_y$ and $\lambda$ do not appear in this equation. On the other hand, it is necessary to mention that $\Omega_x$ includes all the extra corrections that may cause significant deviation from standard cosmology. Therefore, the time evolution of this parameter is of key importance here. Using equation \eqref{E12} and \eqref{E5}-\eqref{E6} we find
\begin{equation}
	\frac{2\dot{H}}{3H^2}=-\Big(\Omega_m+\frac{4}{3}\Omega_r+\Omega_x+\Omega_y\Big)
\end{equation}
we use this equation and the following continuity equations
\begin{equation}
	\begin{split}
		&\dot{\rho}_m+3 H\rho_m=0\\&
		\dot{\rho}_d+3 H\rho_d=0\\&
		\dot{\rho}_r+4 H\rho_r=0
	\end{split}
\end{equation}
to obtain the following dynamical system equations
\begin{equation}\label{E13}
	\Omega_m'=3\Omega_m\Big(\Omega_m+\frac{4}{3}\Omega_r+\Omega_x+\Omega_y-1\Big)
\end{equation}
\begin{equation}
	\Omega_r'=3\Omega_r\Big(\Omega_m+\frac{4}{3}\Omega_r+\Omega_x+\Omega_y-\frac{4}{3}\Big)
\end{equation}
\begin{equation}
	\Omega_x'=3\Omega_x\Big(\Omega_m+\frac{4}{3}\Omega_r+\Omega_x+\Omega_y-1\Big)-3\Omega_y
\end{equation}
\begin{equation}\label{E20}
	\Omega_y'=3\Omega_y\Big(\Omega_m+\frac{4}{3}\Omega_r+\Omega_x+\Omega_y-\lambda\Big)
\end{equation}
\begin{equation}\label{E14}
	\lambda'=3\lambda\Big(\lambda-2+\Gamma(\lambda)\Big)=3\lambda f(\lambda)
\end{equation}
The prime sign stands for derivative with respect to $\ln a$. We do not write an equation for $\Omega_{\Lambda}$ as it is not an independent variable. So the cosmology of RMOND deals with a five-dimensional phase space $(\Omega_m,\Omega_r,\Omega_x,\Omega_y,\lambda)$. Notice that the function $\Gamma$ is defined as
\begin{equation}
	\Gamma=\frac{\mathcal{K}_{Q} \mathcal{K}_{QQQ}}{\mathcal{K}_{QQ}^2}
\end{equation}
where $\mathcal{K}_{QQQ}=\frac{d^3\mathcal{K}}{dQ^3}$. $\lambda$ and $\Gamma$ are functions of $Q$. Therefore, in principle one may eliminate $Q$ to obtain $\Gamma$ as a function of $\lambda$. Before moving on to find the fixed points, let's express the deceleration parameter $q=-\ddot{a}a/\dot{a}^2$ and the effective equation of state parameter $\omega_{\text{eff}}$ in terms of the dynamical system variables:
\begin{equation}\label{F1}
	q(t)=-1-\frac{\dot{H}}{H^2}=-1+\frac{3}{2}\Big(\Omega_m+\frac{4}{3}\Omega_r+\Omega_x+\Omega_y\Big)
\end{equation}
\begin{equation}\label{F2}
	\omega_{\text{eff}}=\frac{2q-1}{3}=-1+\Omega_m+\frac{4}{3}\Omega_r+\Omega_x+\Omega_y
\end{equation}
\subsection{Fixed points:}
Now let us find the fixed points associated with the system. To do so, the right-hand side of the dynamical system equations should be set to zero. The roots of the right-hand side of the $\lambda'$ equation, irrespective of the exact form of $\mathcal{K}$, are shown by $\lambda_*$. It turns out that for each root $\lambda_*$, there are four fixed points labeled as $\mathcal{P}_i$, $i=1,...,4$. These points and their properties are shown in Table \ref{tab1}. In the following, we discuss them in more detail.
\begin{table}[]
	\begin{center}
		\vspace{4pt}
		\begin{tabular}{c|c|c|c}
			\hline \hline 
			Lable & ($\Omega_m$,$\Omega_r$, $\Omega_x$, $\Omega_y$, ${\lambda}$) & $\omega_{\text{eff}}$ & $q$ \\  [1ex]
			\hline 
			$\mathcal{P}_1$ & (0, $1$, 0, 0, $\lambda_{\ast}$) & $\frac{1}{3}$ & $1$ \\ [1ex]
			$\mathcal{P}_2$ & ($1-\Omega_x$, 0, $\Omega_x$, 0, $\lambda_{\ast}$) & 0 & $\frac{1}{2}$ \\ [1ex]
			$\mathcal{P}_3$ & (0, 0, 0, 0, $\lambda_{\ast}$) & $-1$ & $-1$  \\ [1ex]
			$\mathcal{P}_4$ & (0, 0, 1, $\lambda_{\ast}-1$, $\lambda_{\ast}$) & $\lambda_{\ast}-1$ & $-1+\frac{3\lambda_{\ast}}{2}$ \\ [1ex]
			\hline
		\end{tabular}
		\caption{ \label{tab1}The fixed points and their characteristics.}
	\end{center}
\end{table}
\subsubsection{$\mathcal{P}_1$: radiation-dominated phase}
The effective equation of state parameter, in this case, is $\omega_{\text{eff}}=1/3$, and the relativistic matter dominates the cosmic soup. On the other hand, as expected, this is an unstable phase. To be specific, let us express our dynamical systems equations \eqref{E13}-\eqref{E14} as $\textbf{x}_i'=f_i(\textbf{x})$ where $\textbf{x}=(\Omega_m, \Omega_r, \Omega_x, \Omega_y, \lambda)$. Then the linear stability matrix components are given by $\mathcal{M}_{ij}=\partial f_i/\partial x_j$. The eigenvalues associated with each fixed point specify the stability of the point. Since our phase space is five-dimensional, there are five eigenvalues $\mathcal{E}_i$ for each point. The eigenvalues and their stability are summarized in Table \ref{tab2}. 

 In this phase the scale factor grows as $a(t)\propto t^{1/2}$.  Therefore, there is a standard radiation-dominated phase in the thermal history of RMOND provided that $\lambda f(\lambda)=0$ at $\lambda_*$.

\begin{table}[]
	\begin{center}
		\vspace{4pt}
		\begin{tabular}{c|c|c}
			\hline \hline 
			Lable & ($\mathcal{E}_1$, $\mathcal{E}_2$, $\mathcal{E}_3$, $\mathcal{E}_4$, $\mathcal{E}_5$) & Stability \\  [1ex]
			\hline 
			$\mathcal{P}_1$ & ($4$, $1$, $1$, $4-3\lambda_*$, $\mathcal{A}$) & Unstable \\ [1ex]
			$\mathcal{P}_2$ & ($3$, $-1$, $0$, $3(1-\lambda_*)$, $\mathcal{A}$) & Unstable \\ [1ex]
			$\mathcal{P}_3$ &  ($-4$, $-3$, $-3$, $-3\lambda_*$, $\mathcal{A}$) & Stable  \\ [1ex]
			$\mathcal{P}_4$ &  ($-3(1-\lambda_*)$, $-3(1-\lambda_*)$, $3\lambda_*$, $-4+3\lambda_*$, $\mathcal{A}$) & Unstable if $\lambda_{\ast}>0$ \\ [1ex]
			\hline
		\end{tabular}
		\caption{ \label{tab2} Eigenvalues of the Jacobian matrix and their stability for the fixed points. $\mathcal{A}$ is defined as $3\Big(\Gamma(\lambda_*)-2+\lambda_*(2+\frac{d\Gamma(\lambda_*)}{d\lambda})\Big)$.}
	\end{center}
\end{table}

\subsubsection{$\mathcal{P}_2$: matter-dominated phase}

$\mathcal{P}_2$ is a fixed line since $\Omega_x$ takes arbitrary values. Since $\omega_{\text{eff}}$ is zero, the scale factor varies as $a(t)\propto t^{2/3}$. This is the same behavior as the matter-dominated phase in $\Lambda$CDM. There are two specific points on this line: 1) the point $\Omega_x=0$ and $\Omega_m=1$ on this fixed line correspond to the standard matter-dominated phase. In other words, despite the existence of extra terms in the effective density $\hat{\rho}$, there is a standard matter-dominated phase in RMOND. 2) The point $\Omega_x=1$ and $\Omega_m=0$: although the normal non-relativistic matter does not have any contributions to the cosmic evolution at this phase, the extra terms combined in $\Omega_x$ still behave like a matter component with $\omega_{\text{eff}}=0$.

For the fixed line $\mathcal{P}_2$, the eigenvalue $\mathcal{E}_1$ is positive, and it is enough to guarantee that all the points on this fixed line are unstable.

\subsubsection{$\mathcal{P}_3$: de Sitter phase}
This fixed point corresponds to $\Omega_{\Lambda}=1$, $\omega_{\text{eff}}=-1$ and $q=-1$.
This point is stable if $\lambda_*>0$ and $\mathcal{A}<0$. If so, then $\mathcal{P}_3$ is a standard de Sitter phase necessary to explain the cosmic speed up. 

\subsubsection{$\mathcal{P}_4$: $\Omega_x$ dominated phase:}
This fixed point does not exist in the standard cosmology. Depending on the value of $\lambda_*$, this fixed point describes an accelerated or decelerated expansion. This fixed point is of special importance in the sense the extra terms proportional to $a^{-n}$ with $n>3$ in the effective energy density control the cosmic evolution. So it is important to ensure that these terms do not cause serious deviations from the standard cosmology. Since such a phase does not exist in the standard cosmology, it is natural to expect that the cosmic evolution should not stay too long in this phase. This directly means that this fixed point should be unstable. Otherwise, there would be a false phase that the cosmic evolution falls into it and stays there forever.

It should be noted that for $\lambda_*>0$, the third eigenvalue is positive $\mathcal{E}_3>0$, and the point is unstable.  Fortunately the condition $\lambda_*>0$ is also necessary for having a stable de Sitter phase. In other word, at the same time, we have a viable de Sitter phase and an unstable $\Omega_x$ dominated phase.
\subsection{Different potentials:}
In our analysis in the previous subsection, we did not choose any specific potential $\mathcal{K}(Q)$. Now, let us investigate all the potentials proposed in RMOND. We start with the simple quadratic potential $\mathcal{K}_1(Q)$.
\subsubsection{The quadratic potential $\mathcal{K}_1(Q)$}
In this case it is easy to show that 
\begin{equation}
	\lambda(Q)=2, \,\,\,\,\,\,\,\,\,\,\,\,\, \Gamma(Q)=0
\end{equation}
this means that for this specific case, the phase space is four-dimensional since $\lambda$ is constant. Therefore we have four eigenvalues for each point. One may simply ignore the eigenvalue $\mathcal{E}_5$. 

This potential possesses a true unstable matter-dominated phase. This is the case also for the radiation-dominated phase. The late time de Sitter solution is stable as expected. Although $\mathcal{P}_4$ is unstable, it appears in a wrong place and causes violent deviations from the standard cosmology.

\begin{figure}[tbp]
	\centering
	\includegraphics[width=.6\textwidth]{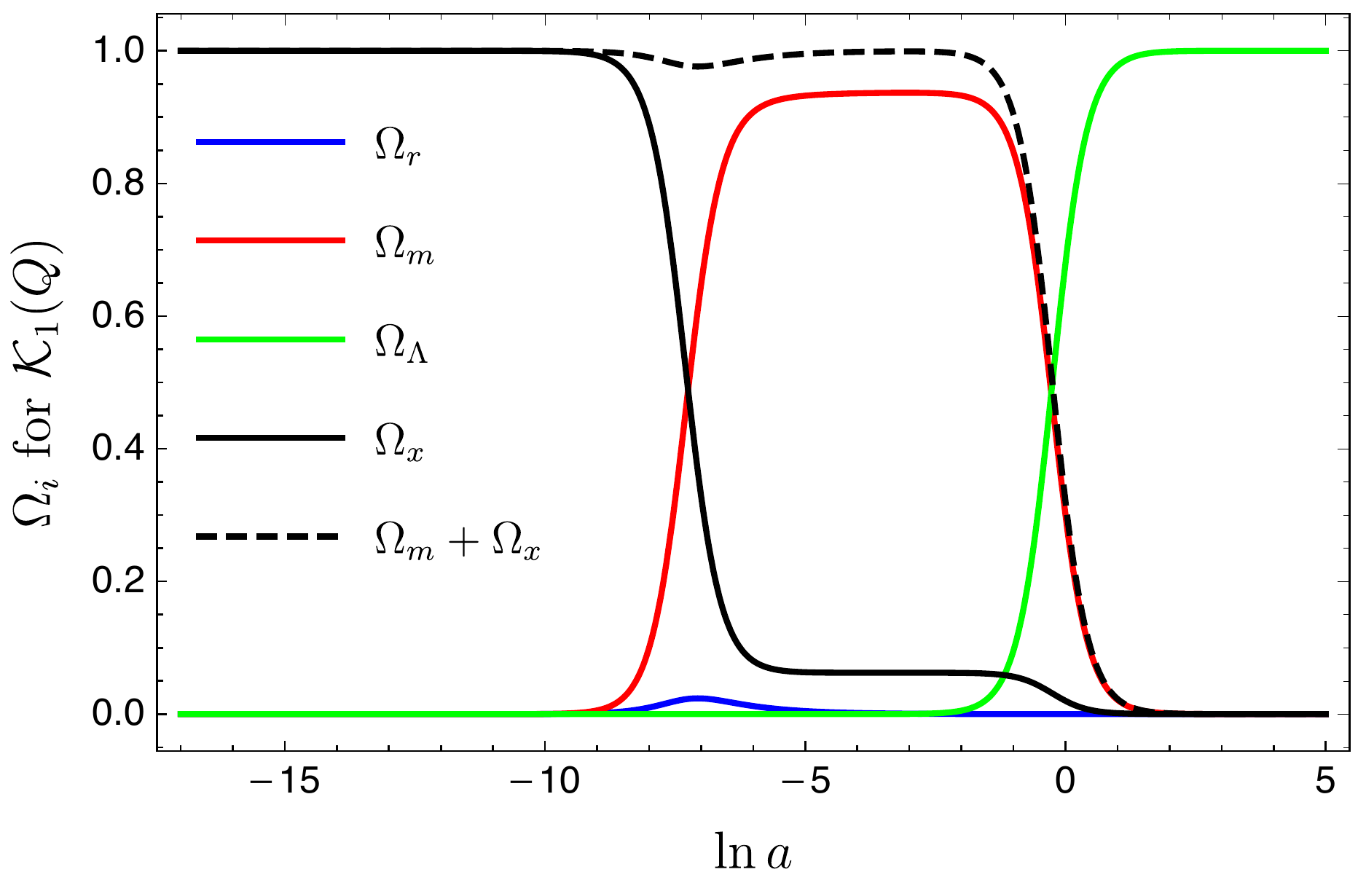}
	\includegraphics[width=.6\textwidth]{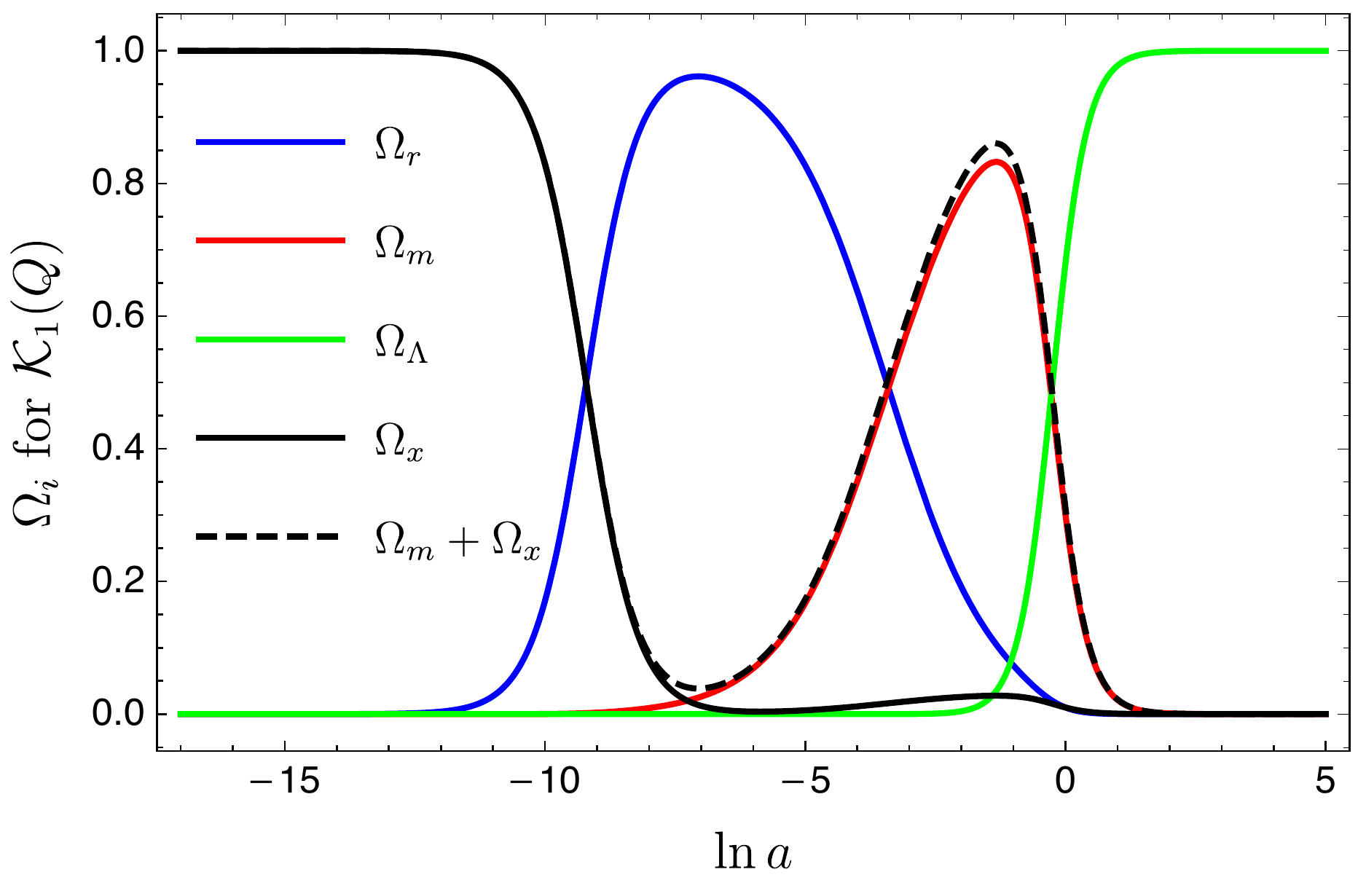}
	\caption{\label{Quadra_1} The evolution of the cosmic density parameters in the quadratic potential $\mathcal{K}_1(Q)$. In the top panel the initial conditions are set at $\ln a=0$ as $\Omega_m=0.3$, $\Omega_r=10^{-5}$, $\Omega_y=10^{-10}$ and $\Omega_{\Lambda}=0.68$. In the bottom panel everything is the same as in the top panel but $\Omega_r=10^{-2}$.}
\end{figure}

The time evolution of the cosmic densities for two specific choices of initial conditions has been shown in Fig.~\ref{Quadra_1}. Each panel corresponds to a specific trajectory in the phase space. In the top panel, we have chosen the current observational values of the cosmic densities as the initial conditions. The evolution starts with the decelerating phase $\mathcal{P}_4$ with $\Omega_x=1$ and $\omega_{\text{eff}}=1$, and then continues to the unstable matter-dominated phase $\mathcal{P}_2$.
Finally, it reaches the de Sitter phase $\mathcal{P}_3$. The radiation-dominated point $\mathcal{P}_1$ is never realized with these specific initial conditions. Therefore, clearly, this is a wrong evolution for the cosmic background. 

To capture the radiation-dominated phase, we use a much higher value for the current magnitude of $\Omega_r$. On the other hand, we keep all the other initial conditions unchanged. The evolution of the cosmic densities is shown in the bottom panel of Fig.~\ref{Quadra_1}. In this case, the evolution starts with $\mathcal{P}_4$, then enters the radiation-dominated phase $\mathcal{P}_1$. Then the trajectory in the phase space gets close to $\mathcal{P}_2$ and eventually falls into the late time accelerated fixed point. However, the existence of $\mathcal{P}_4$ at the early universe clearly signals a wrong evolution.

To summarize, let us emphasize that the quadratic potential $\mathcal{K}_1(Q)$ is not viable. This fact has already been reported in \cite{Skordis:2020eui} based on different analyses. Notice that with this potential, the total energy density falls as $A/a^3+B/a^4+C/a^6$. Each term dominates in a specific period of time. Therefore, it is natural to expect an extra phase associated with the last term. As the dynamical system analysis proved, the existence of this extra phase destroys the validity of the quadratic model. 

\subsubsection{The Higgs-like potential $\mathcal{K}_2(Q)$}
\begin{figure}[tbp]
	\centering
	\includegraphics[width=.6\textwidth]{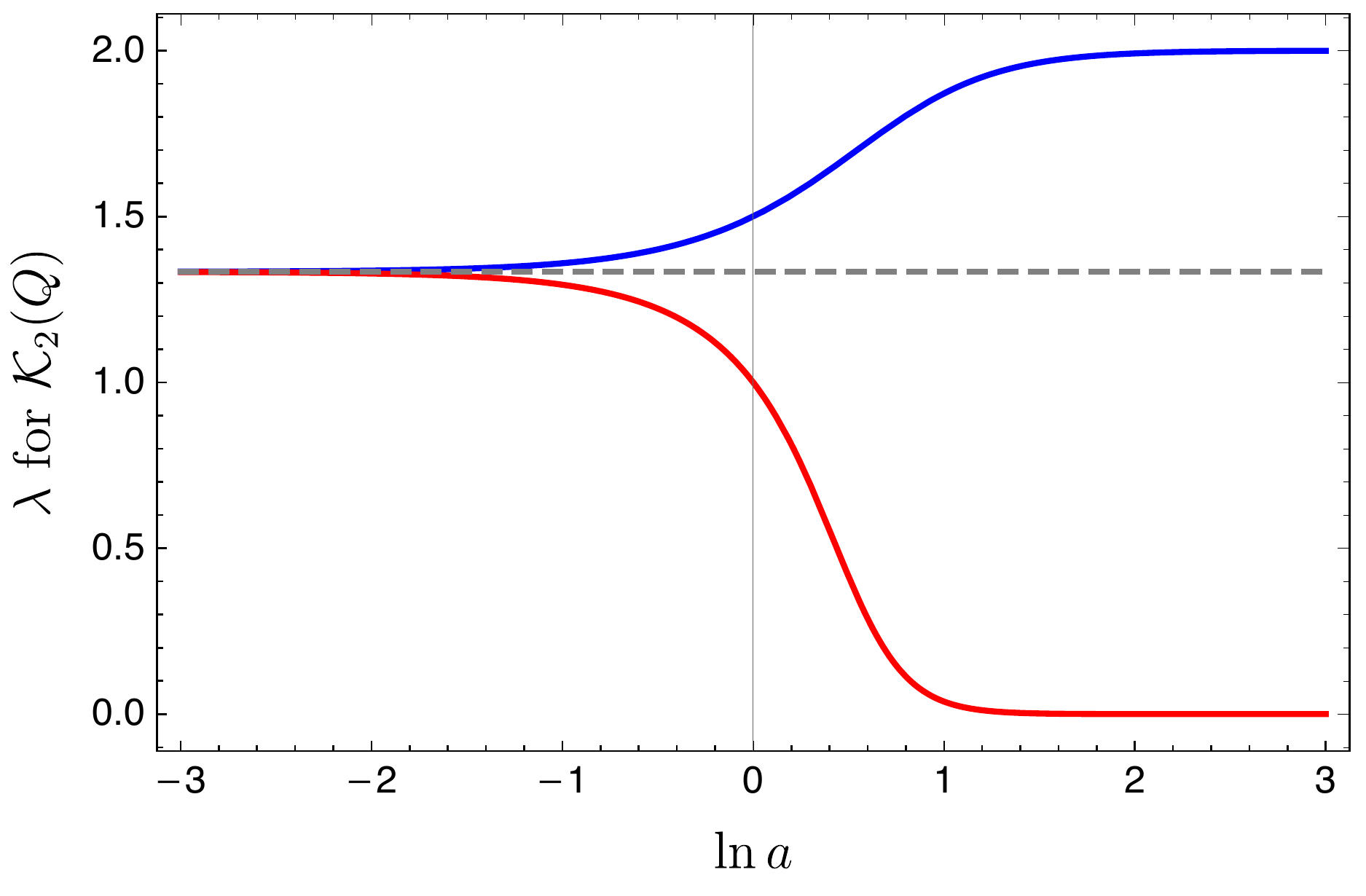}
	\caption{\label{lambda_higgs} The evolution of $\lambda$ for the Higgs-like potential $\mathcal{K}_2(Q)$. In this model, $\lambda$ starts at $\lambda=4/3$, and then two separate branches, shown by red and blue curves, are possible. The dashed gray line indicates the line $\lambda=\frac{4}{3}$. }
\end{figure}

In this case it is easy to show that
\begin{equation}
	\lambda(Q)=\frac{4Q^2}{3Q^2-Q_0^2}, \,\,\,\,\,\,\,\,\,\,\,\,\, \Gamma(\lambda)=\frac{3}{4}\lambda(2-\lambda)
\end{equation}
$Q_0$ indicates the minimum of the potential at the late time de Sitter phase. At the late time by $Q\rightarrow Q_0$, we find $\lambda_{\infty}=2$ and $\Gamma_{\infty}=0$. The equation \eqref{E14} can be analytically integrated as
\begin{equation}\label{E15}
	\Big(\frac{\mid4-3 \lambda\mid}{\mid4-3 \lambda_0\mid}\Big)^3 \times\Big(\frac{\mid2- \lambda_0\mid}{\mid2- \lambda\mid}\Big)^2 \times\frac{\mid\lambda_0\mid}{\mid\lambda\mid}=a^6
\end{equation}
where $\lambda_0$ is the current value of $\lambda$, namely $\lambda(0)=\lambda_0$. It should be noted that the current value of the scale factor is scaled to unity throughout this paper. By taking the limit of \eqref{E15} at $a\rightarrow 0$ we find that $\lambda$ always starts with $\lambda=\frac{4}{3}$. There are two general degenerate branches for the time evolution of $\lambda$ shown in Fig.~\ref{lambda_higgs}. By degenerate, we mean that although the evolution of $\lambda$ is different, the other phase variables are the same in both branches. They can be classified by the value of $\lambda_0$. By looking at the right-hand side of \eqref{E14}, and assuming that $0\leq\lambda\leq2$ it is clear that $\lambda_0'>0$ if $\frac{4}{3}<\lambda_0<2$. In this case, the blue curve is realized. On the other hand, $\lambda_0'<0$ if $0<\lambda_0<\frac{4}{3}$ and the red curve is achieved. The blue branch asymptotically reaches the value $\lambda=2$, and accordingly, the red curve asymptotically reaches $\lambda=0$. Interestingly, each asymptotic value corresponds to a set of fixed points. More specifically, it is easy to show that there are three roots for $\lambda f(\lambda)$ as $\lambda_*=0,\frac{4}{3}, 2$. As already mentioned, for each value of $\lambda_*$, in principle, there are four fixed points. Lets us label the 12 fixed points/lines as $\mathcal{P}_i^{[0]}$, $\mathcal{P}_i^{[4/3]}$ and $\mathcal{P}_i^{[2]}$ for $i=1,2,3,4$.

\begin{figure}[tbp]
	\centering
	\includegraphics[width=.6\textwidth]{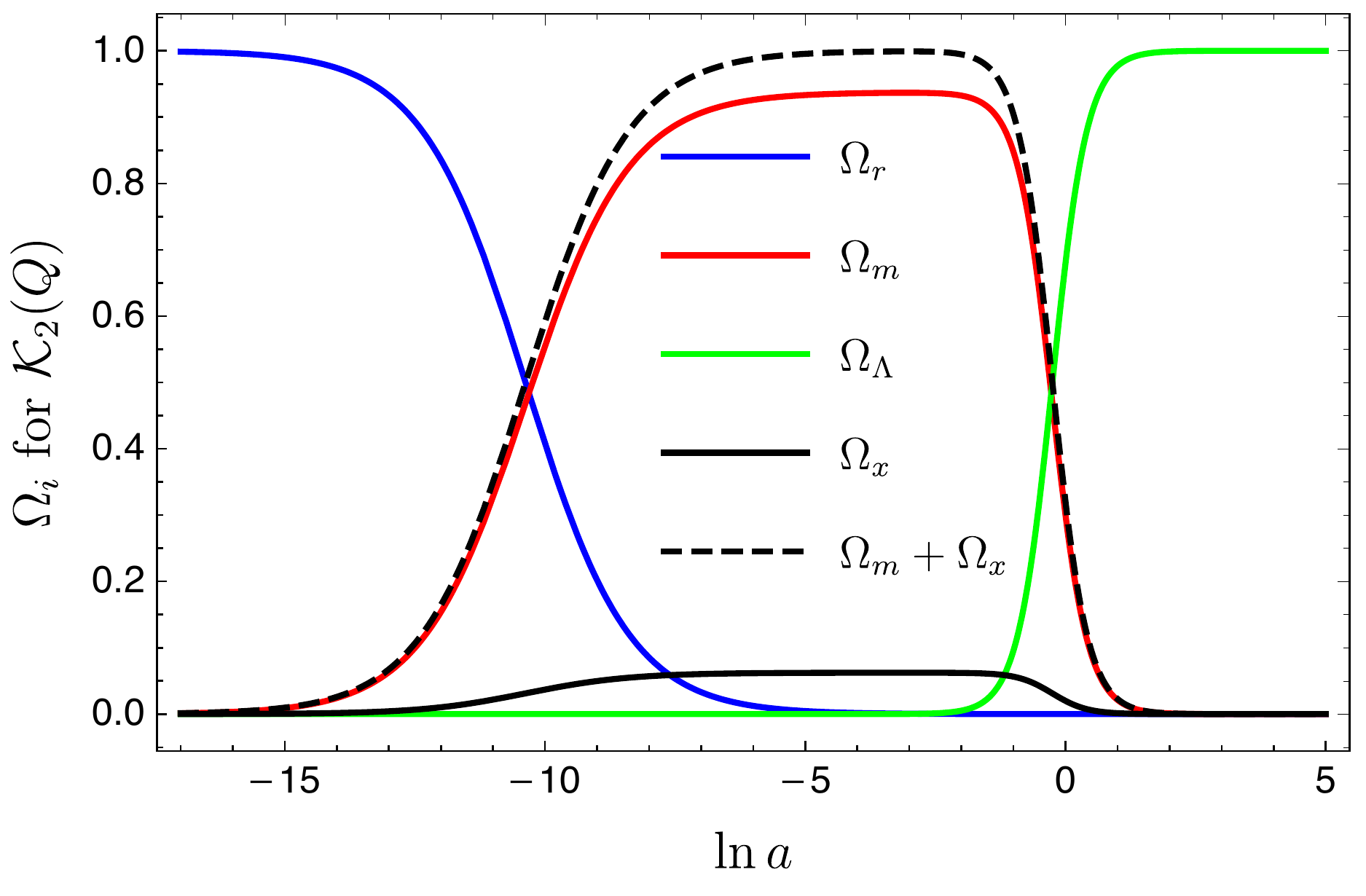}
	\caption{\label{Higgs_2} The evolution of the cosmic density parameters in the Higgs-like potential $\mathcal{K}_2(Q)$. The initial conditions are set at $\ln a=0$ as $\Omega_m=0.3$, $\Omega_r=10^{-5}$, $\Omega_y=10^{-10}$, $\lambda_0=1.8$ and $\Omega_{\Lambda}=0.68$.}
\end{figure}

Now let us pick the same initial conditions used in the quadratic model. The result has been shown in Fig.~\ref{Higgs_2}. The evolution starts with the standard radiation-dominated phase $\mathcal{P}_1^{[4/3]}$ then enters the standard matter-dominated phase $\mathcal{P}_2^{[4/3]}$. In the end, the evolution falls into the late time de Sitter phase $\mathcal{P}_3^{[2]}$ (or $\mathcal{P}_3^{[0]}$). As mentioned before this phase is stable if $\lambda_*>0$ and $\mathcal{A}<0$. It is easy to show that this potential has a standard accelerated phase because both fixed points $\mathcal{P}_3^{[0]}$ and $\mathcal{P}_3^{[2]}$ have a negative value for $\mathcal{A}$. It should be noted that the evolution of the density parameters, shown in  Fig.~\ref{Higgs_2}, is almost insensitive to the value of $\lambda_0$. Notice that trajectory in the phase space does not pass the nonstandard extra points $\mathcal{P}_4^{[0]}$, $\mathcal{P}_4^{[4/3]}$ and $\mathcal{P}_4^{[2]}$. It should be emphasized that since the evolution starts with $\lambda_*=4/3$, unlike in the quadratic model, the points $\mathcal{P}_4^{[0]}$ and  $\mathcal{P}_4^{[2]}$ do not appear in the early universe. 

It is interesting to mention that, in the Higgs-like potential, the point $\mathcal{P}_4^{[4/3]}$ is characterized by $\omega_{\text{eff}}=1/3$. In other words, this point behaves like a radiation-dominated phase. However, notice that this is not a standard radiation-dominated phase in the sense that the relativistic matter has no contribution. This phase does not appear in the solution presented in Fig.~\ref{Higgs_2}. 

\begin{figure}[tbp]
	\centering
	\includegraphics[width=.5\textwidth]{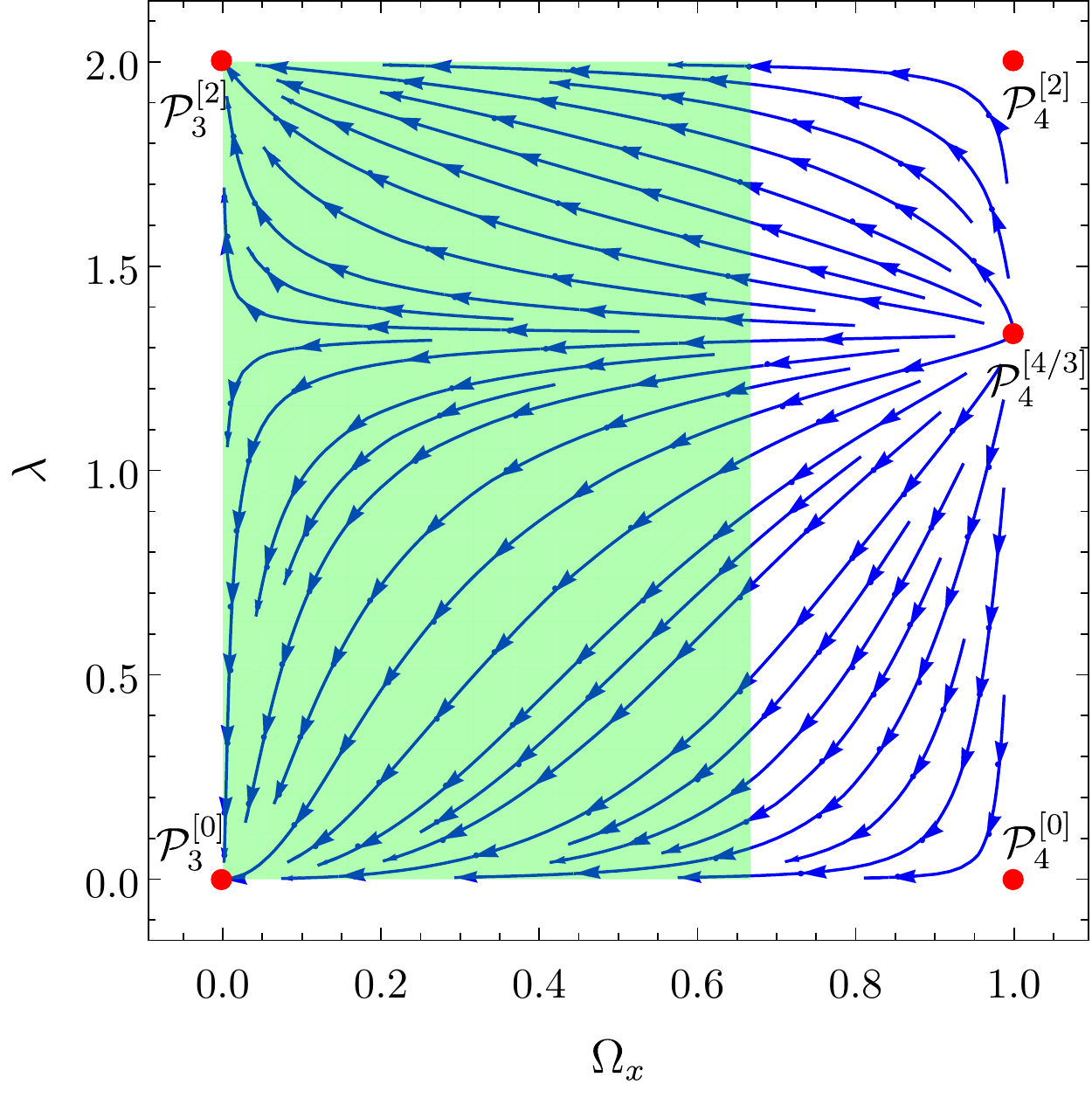}
	\caption{\label{Higgs_3} The projection of phase space on $\Omega_x-\lambda$ plane for the Higgs-like potential. $\mathcal{P}_4^{[4/3]}$, $\mathcal{P}_4^{[0]}$ and $\mathcal{P}_4^{[2]}$ are unstable fixed points. The stable fixed points $\mathcal{P}_3^{[0]}$ and $\mathcal{P}_3^{[2]}$ indicate the de Sitter phase. The green (shaded) region shows the accelerated expansion phase.}
\end{figure}

Figure~\ref{Higgs_3} illustrates the projected phase space for autonomous equations \eqref{E13}-\eqref{E14} on the $\Omega_x-\lambda$ plane. For all values of $\lambda_*$, the points $\mathcal{P}_4^{[\lambda_*]}$ are unstable as expected. On the other hand, the points $\mathcal{P}_3^{[0]}$ and $\mathcal{P}_3^{[2]}$ are stable. These stable points indicate the late time de Sitter phase. The green area illustrates the region where the cosmic expansion is accelerated, i.e., $\ddot{a}>0$.

To summarize, the problem with the quadratic model is resolved in the Higgs-like model. This model gives a viable sequence of fixed points resembling the standard cosmological model, provided that an appropriate set of initial conditions is used.

\subsubsection{The "Cosh" potential $\mathcal{K}_3(Q)$}
For this potential we have
\begin{equation}
	\lambda(Q)=1+\text{sech}\mathcal{Z}, \,\,\,\,\,\,\,\,\,\,\,\,\, \Gamma(\lambda)=\lambda(2-\lambda)
\end{equation}
The equation \eqref{E14} can be integrated to give
\begin{equation}
	\lambda(a)=1\pm \frac{a^3}{\sqrt{a^6+\beta}}
\end{equation}
where $\beta=-1+(1-\lambda_0)^{-2}$ and $\lambda_0$ is the current value of $\lambda$. Therefore $\lambda$ starts with $\lambda=1$ at early times and then asymptotically reaches $0$ or $2$. In other words, in this case, we have $\lambda_* =0, 1, 2$. Let's indicate the corresponding fixed points/lines by $\mathcal{P}_i^{[0]}$, $\mathcal{P}_i^{[1]}$ and $\mathcal{P}_i^{[2]}$. Since $\lambda$ starts from $\lambda=1$ at the early universe, the unusual points $\mathcal{P}_4^{[0]}$ and $\mathcal{P}_4^{[2]}$ do not show up. On the other hand, in this case, the fixed point $\mathcal{P}_4^{[1]}$ lies on the fixed line $\mathcal{P}_2^{[1]}$ and corresponds to a matter-dominated phase where $\omega_{\text{eff}}=0$. It is necessary to mention that although the scale factor varies as $a(t)\propto t^{2/3}$,  this fixed point is not standard in the sense that the normal non-relativistic matter does not contribute to this phase.

\begin{figure}[tbp]
	\centering
	\includegraphics[width=.6\textwidth]{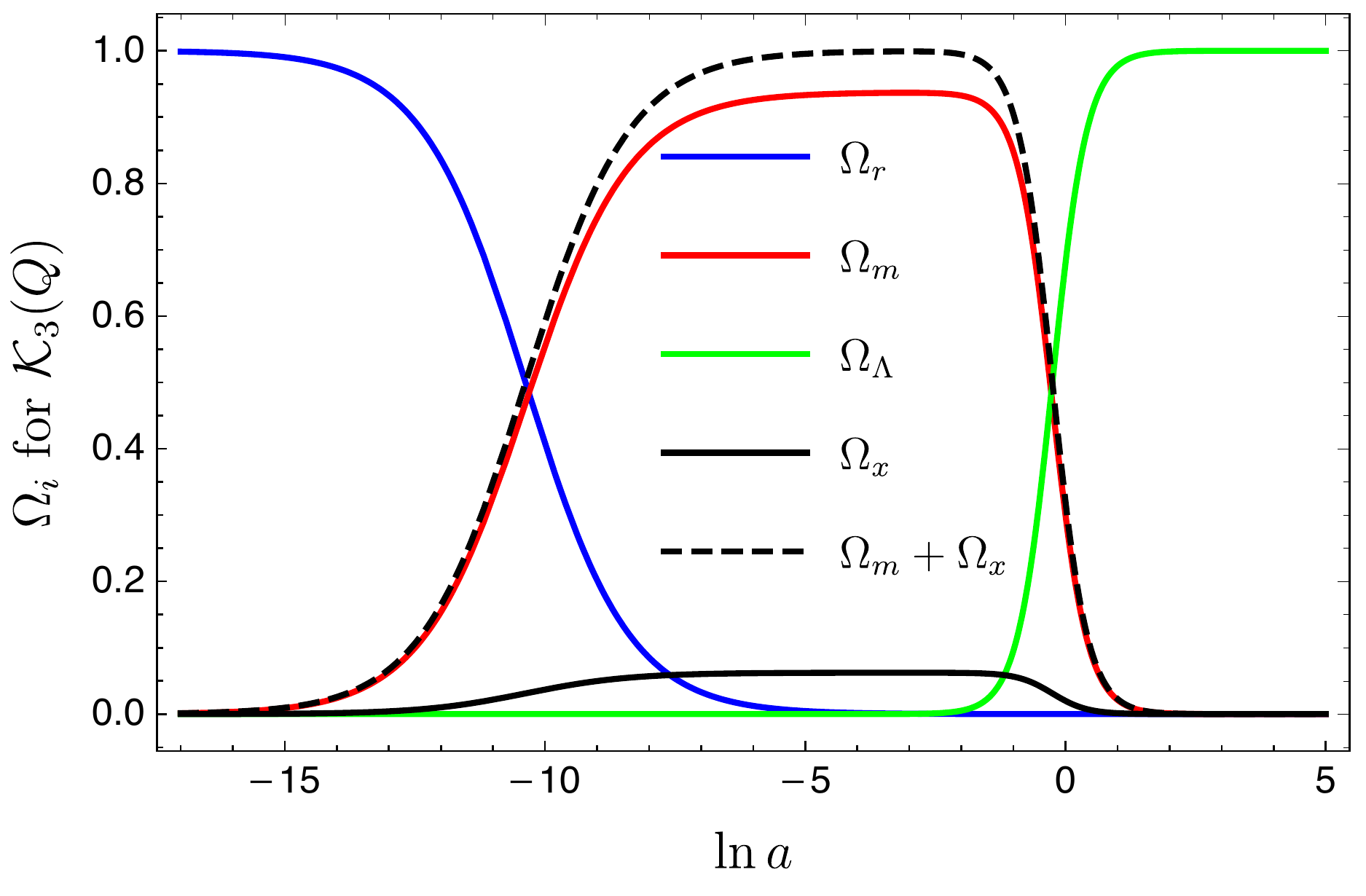}
	\caption{\label{Cosh_1} The evolution of the cosmic density parameters in the case of "Cosh" potential $\mathcal{K}_3(Q)$. The initial conditions are set at $\ln a=0$ as $\Omega_m=0.3$, $\Omega_r=10^{-5}$, $\Omega_y=10^{-10}$, $\lambda_0=1.8$ and $\Omega_{\Lambda}=0.68$.}
\end{figure}

In Fig.~\ref{Cosh_1} we have illustrated the evolution of the cosmic density parameters. The same initial conditions as in the Higgs-like potential are implemented. It is difficult to discriminate the evolution of the density parameters in these models. Notice that the trajectory in the phase space pass the following points/line $\mathcal{P}_1^{[1]}$, $\mathcal{P}_2^{[1]}$ and $\mathcal{P}_3^{[2]}$ (or $\mathcal{P}_3^{[0]}$). Since $\mathcal{A}$ is negative, the stability of the late time de Sitter phase, i.e., $\mathcal{P}_3^{[0]}$ and $\mathcal{P}_3^{[2]}$, is guaranteed. The problematic fixed point $\mathcal{P}_4^{[1]}$ does not appear in this solution. 

\begin{figure}[tbp]
	\centering
	\includegraphics[width=0.5\textwidth]{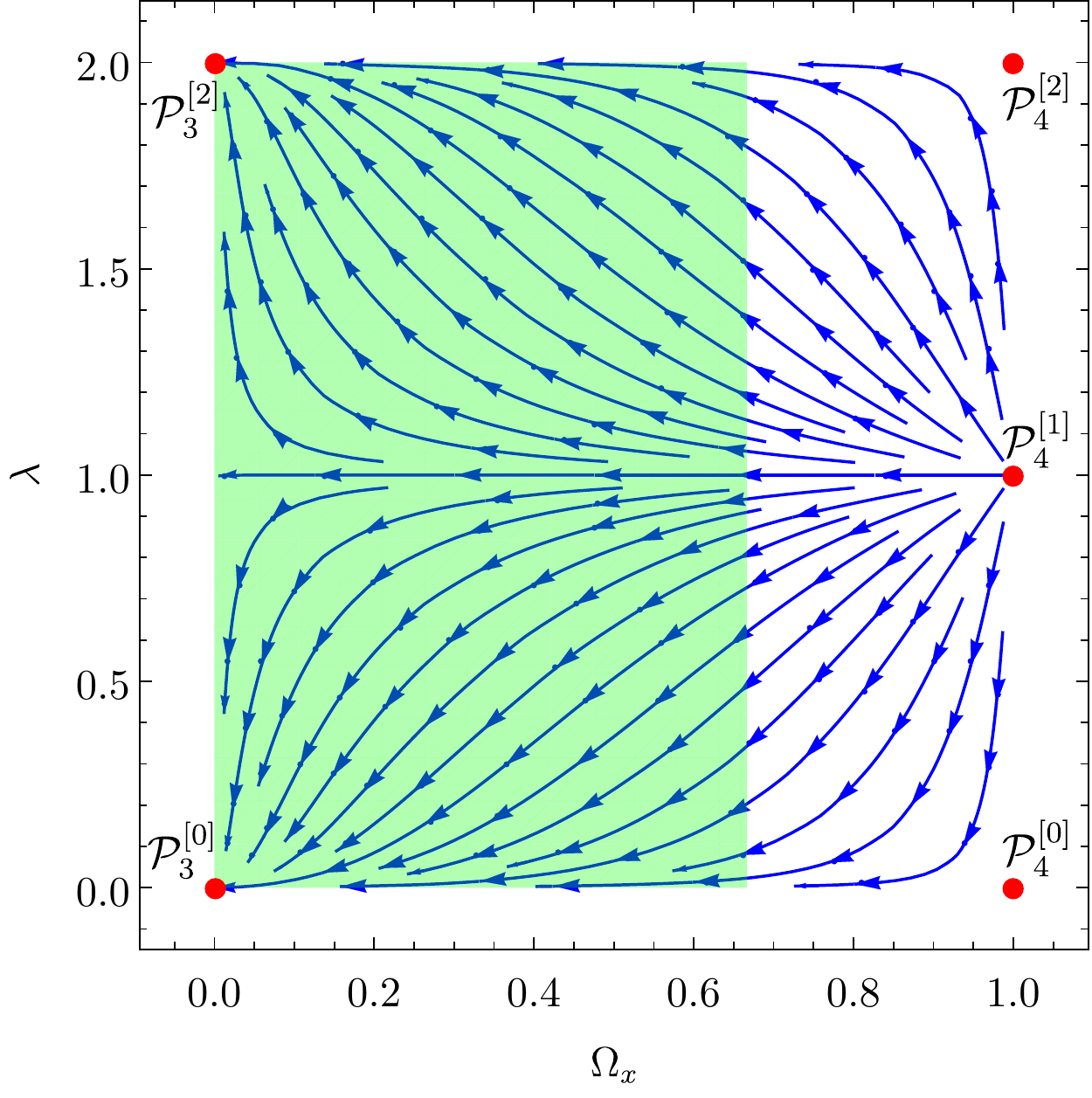}
	\caption{\label{Cosh_3} The projected phase space on $\Omega_x-\lambda$ plane for the "Cosh" potential. There are three unstable points corresponding to the $\Omega_x$ dominated phase, $\mathcal{P}_4^{[1]}$, $\mathcal{P}_4^{[0]}$ and $\mathcal{P}_4^{[2]}$. The de Sitter phase is denoted by $\mathcal{P}_3^{[0]}$ and $\mathcal{P}_3^{[2]}$. The green (shaded) area indicates the accelerated regime. }
\end{figure}

The projected phase space for this potential has been plotted in Fig.~\ref{Cosh_3}. Trajectories with different initial conditions start from $\mathcal{P}_4^{[1]}$ which is a repulsive point, and get close to two saddle points $\mathcal{P}_4^{[0]}$ and $\mathcal{P}_4^{[2]}$. All the trajectories finally enter the stable points $\mathcal{P}_3^{[0]}$ and $\mathcal{P}_3^{[2]}$ that represent the late time de Sitter phase. The green (shaded) region in this figure denotes the area which $\omega_{\text{eff}}<-\frac{1}{3}$ i.e., the accelerated expansion regime.

\subsubsection{The exponential potential $\mathcal{K}_4(Q)$}

Mathematically, the exponential model is more complex than the other models as far as the background cosmology is concerned. In this case, the functions $\lambda$ and $\Gamma$ are
\begin{equation}
	\lambda(Q)=\frac{2 e^{\mathcal{Z}^2}\mathcal{Z}^2}{(e^{\mathcal{Z}^2}-1)(1+2 \mathcal{Z}^2)},\,\,\,\,\,\,\Gamma(Q)=\frac{2\mathcal{Z}^2 (3+2 \mathcal{Z}^2)}{(1+2 \mathcal{Z}^2)^2}
\end{equation}
Although one can easily find $\lambda(\Gamma)$, it is not possible to find an expression for $\Gamma(\lambda)$. Therefore, there would be some difficulties with the numeric solutions of the dynamical system differential equations. In the following, we replace $\lambda$ with a new definition that is suitable for the exponential potential. However, before moving on, let us mention that one may numerically find $\lambda_*$s for our original definition of $\lambda$. To do so, let us plot the right-hand side of \eqref{E14} in terms of $\mathcal{Z}$, namely $\lambda'(\mathcal{Z})$. It turns out that $\lambda'$ vanishes at $\mathcal{Z}=0, \pm1.793,$ and $\pm\infty$. Accordingly, by taking the limit of $\lambda(\mathcal{Z})$ at these values we find $\lambda_*=2, 0.90,$ and $1$. Notice that $\mathcal{P}_4^{[0.9]}$ corresponds to an accelerated phase with $\omega_{\text{eff}}=-0.1$.

\begin{figure}[tbp]
	\centering
	\includegraphics[width=.6\textwidth]{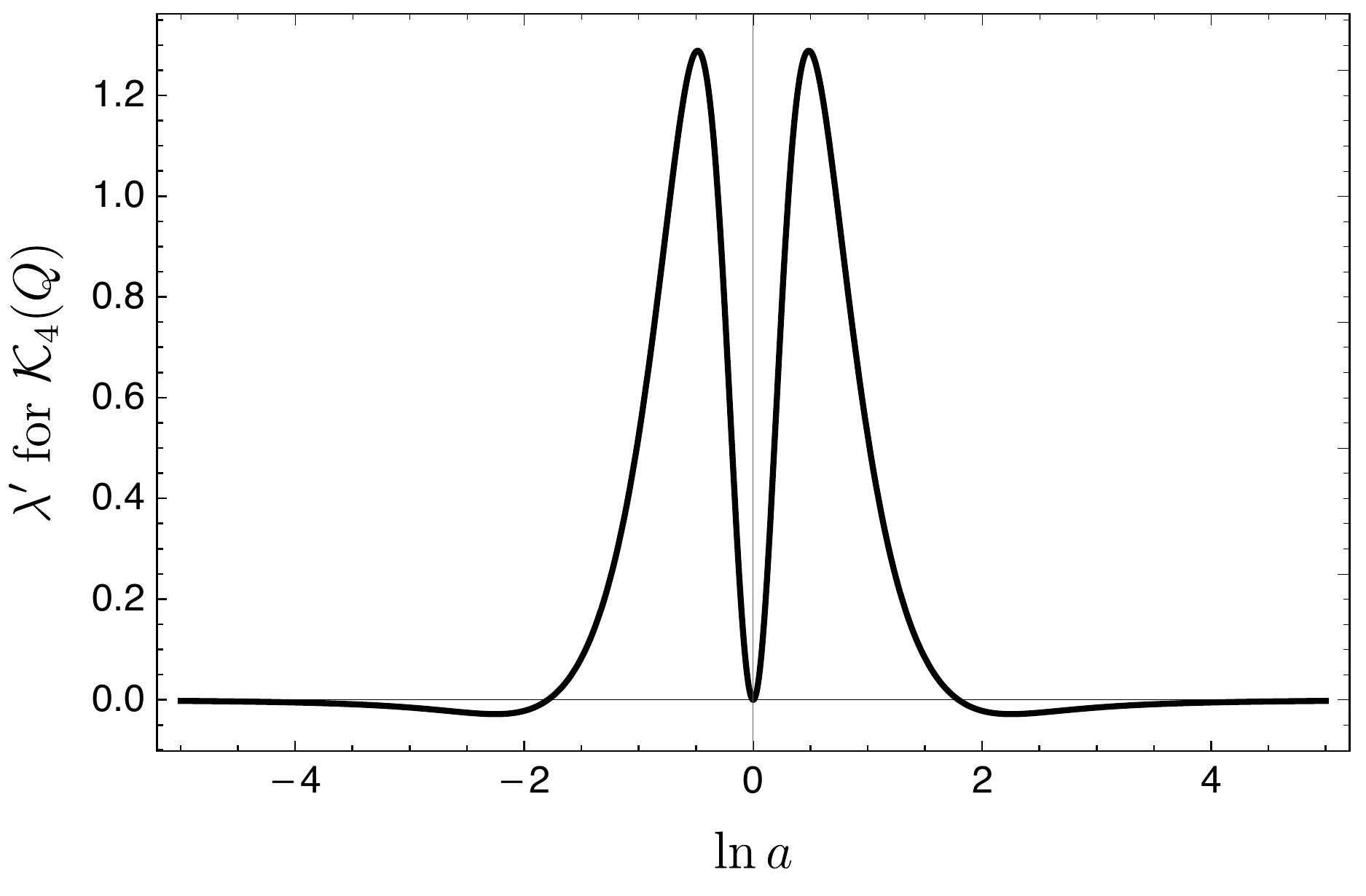}
	\caption{\label{Exp_1} $\lambda'$ in terms of $\ln a$ for the exponential potential $\mathcal{K}_4(Q)$.}
\end{figure}

Our analysis so far is powerful in the sense that the free parameters of the theory do not appear in the calculations. The existence of the viable cosmic epochs is explored independently of the magnitude of the free parameters. This is the case also in GR when the magnitude of the gravitational constant $G$ and the velocity of light do not directly appear in the dynamical system analysis. Now we redefine $\lambda$ for the exponential potential. The only cost is that we will need the magnitude of the free parameters. It proves useful to define $\tilde{\lambda}$ and $\tilde{\Gamma}$ as follows
\begin{equation}\label{E21}
	\begin{split}
		&\tilde{\lambda}=\frac{\mathcal{K}_{Q}}{\mathcal{Z}_0 \mathcal{K}_{QQ}}=\frac{\mathcal{Z}}{1+2\mathcal{Z}^2},\\&\tilde{\Gamma}(\tilde{\lambda})=\frac{\mathcal{Z}_0\mathcal{K}_{QQQ}}{\mathcal{K}_{QQ}}=\frac{6\mathcal{Z}+4\mathcal{Z}^3}{1+2\mathcal{Z}^2}=\frac{1+8\tilde{\lambda}^2-\sqrt{1-8\tilde{\lambda}^2}}{2 \tilde{\lambda}}
	\end{split}
\end{equation}
$\tilde{\lambda}$ satisfies the following equation
\begin{equation}\label{E21}
	\tilde{\lambda}'=-\frac{3}{2}\tilde{\lambda}\Big(1-8\tilde{\lambda}^2+\sqrt{1-8\tilde{\lambda}^2}\Big)
\end{equation}
which in this case, we find three roots $\tilde{\lambda}_*=\pm 1/\sqrt{8}, 0$.
 
\begin{figure}[tbp]
	\centering
	\includegraphics[width=.6\textwidth]{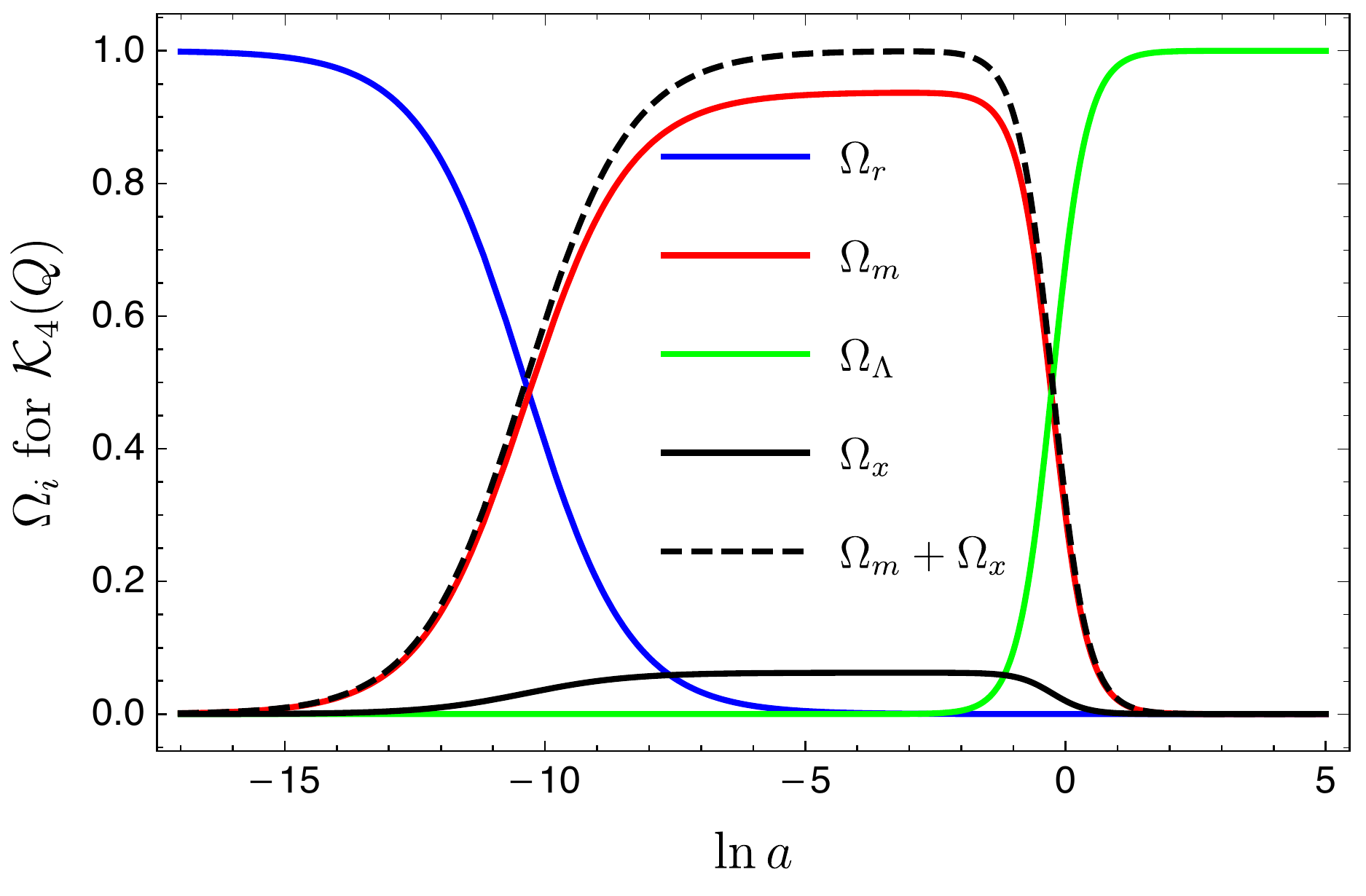}
	\caption{\label{Exp_2} The evolution of the cosmic density parameters in the exponential potential $\mathcal{K}_4(Q)$. The initial conditions are set at $\ln a=0$ as $\Omega_m=0.3$, $\Omega_r=10^{-5}$, $\Omega_y=10^{-10}$, and $\Omega_{\Lambda}=0.68$. On the other hand, $\tilde{\lambda}(-17)=1/\sqrt{8}$ and $Q_0/\mathcal{Z}_0=10^{13}$ as in \cite{Skordis:2020eui}.}
\end{figure}

Except for the equation \eqref{E20}, the other equations of the dynamical system do not change. We need to replace \eqref{E20} with
\begin{equation}\label{E22}
	\Omega_y'=-3\left(\frac{\mathcal{Z}_0}{Q_0}\right)(\Omega_m-\Omega_b)\tilde{\lambda}+3\Omega_y\Big(\Omega_m+\Omega_x+\Omega_y+\frac{4}{3}\Omega_r\Big)
\end{equation}
where $\Omega_b$ is the baryonic cosmic density parameter and satisfies the following equation
\begin{equation}
	\Omega_b'=3\Omega_b\Big(\Omega_m+\Omega_x+\Omega_y+\frac{4}{3}\Omega_r-1\Big)
\end{equation}
Now we have to choose suitable values of $\mathcal{Z}_0$ and $Q_0$ compatible with CMB observations. We pick them from \cite{Skordis:2020eui} as $\mathcal{Z}_0=10^{-17}$ and $Q_0=10^{-4}$. Notice that from \eqref{E21} we have $\mid\tilde{\lambda}\mid\leq1/\sqrt{8}$. Therefore, the first term on the right-hand side of \eqref{E22} is extremely small. So we neglect this term. We have numerically solved the dynamical system equations for the same initial conditions as in the other models. The result is shown in Fig.~\ref{Exp_2}. We see that the background cosmology follows a standard trajectory in the phase space similar to the standard $\Lambda$CDM model.

\subsection{Comparison with $\Lambda$CDM}
\begin{figure}[tbp]
	\centering
	\includegraphics[width=.6\textwidth]{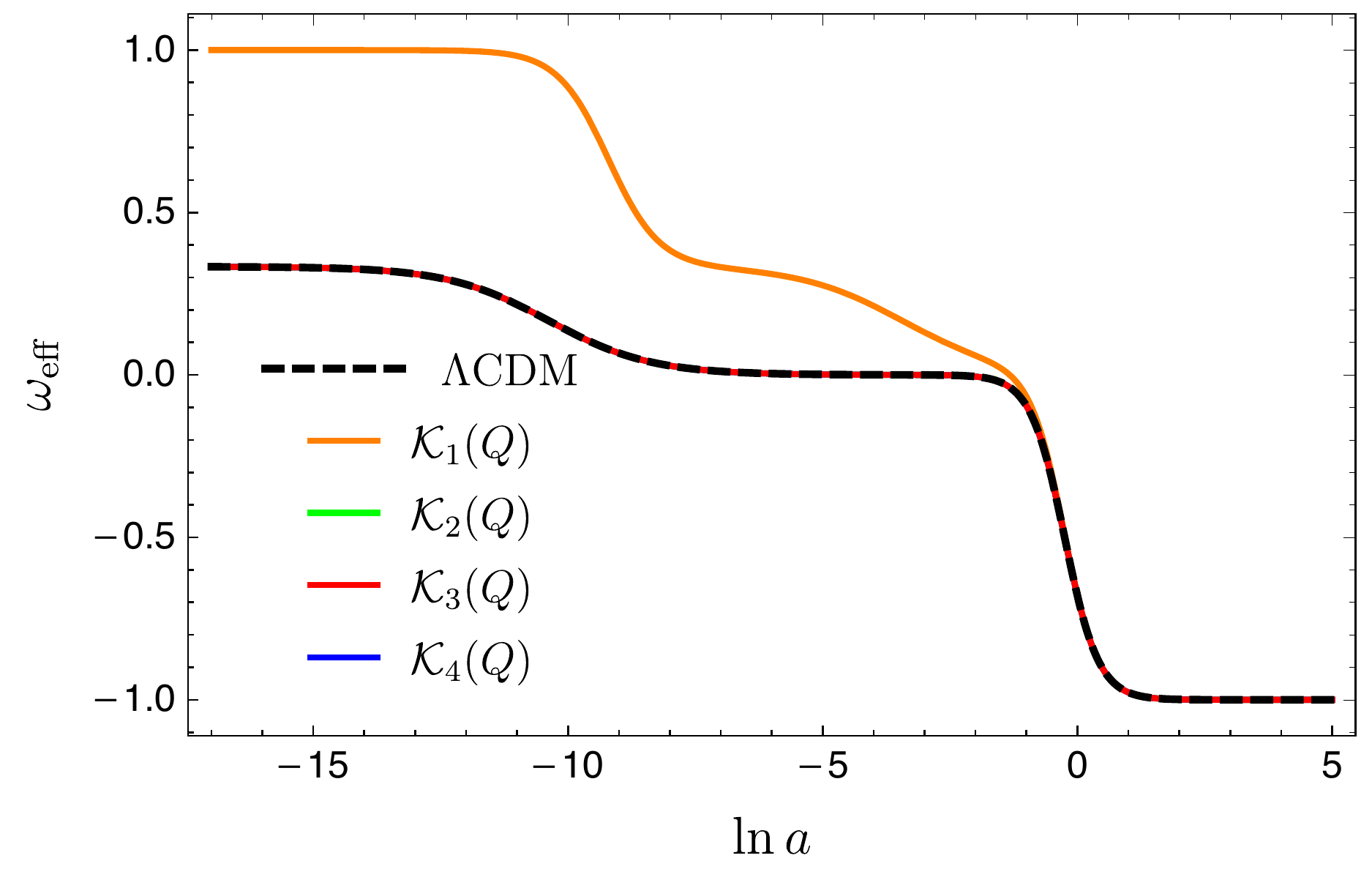}
	\caption{\label{omega} The evolution of the effective equation of state parameter in RMOND models compared with $\Lambda$CDM.}
\end{figure}

We confirmed that the quadratic model $\mathcal{K}_1$ does not work. On the other hand, the other three models give a suitable sequence of fixed points with a true expansion rate. It would be instructive to compare the evolution of density parameters in RMOND with those in standard cosmology. To do so, we compare the effective equation of state parameter $\omega_{\text{eff}}$ in different models. In the case of $\Lambda$CDM we use the initial conditions at $\ln a=0$ as $\Omega_r=10^{-5}$ and $\Omega_{\Lambda}=0.68$. This is the same as that of we used in the RMOND models. Notice that the initial condition on $\Omega_m$ in $\Lambda$CDM should be equal to $\Omega_m+\Omega_x$ at $\ln a=0$ in RMOND models. The evolution of $\omega_{\text{eff}}$ is shown in Fig.~\ref{omega}. The three cosmological epochs, namely the matter, radiation, and dark energy-dominated phases, can be clearly seen as step-like features in $\omega_{\text{eff}}$. As expected, the $\mathcal{K}_1$ model violently deviates from $\Lambda$CDM. On the other hand, the other models cannot be distinguished from $\Lambda$CDM.

\begin{figure}[tbp]
	\centering
	\includegraphics[width=.62\textwidth]{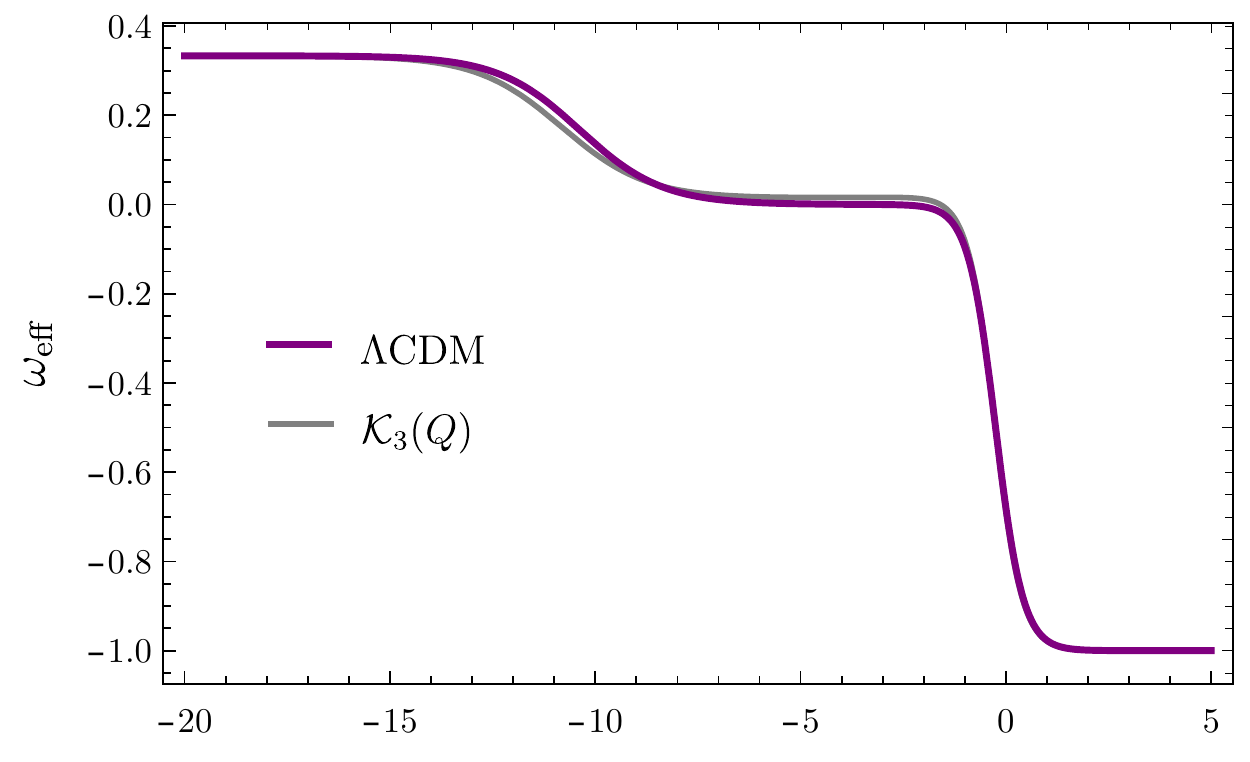}
	\includegraphics[width=.6\textwidth]{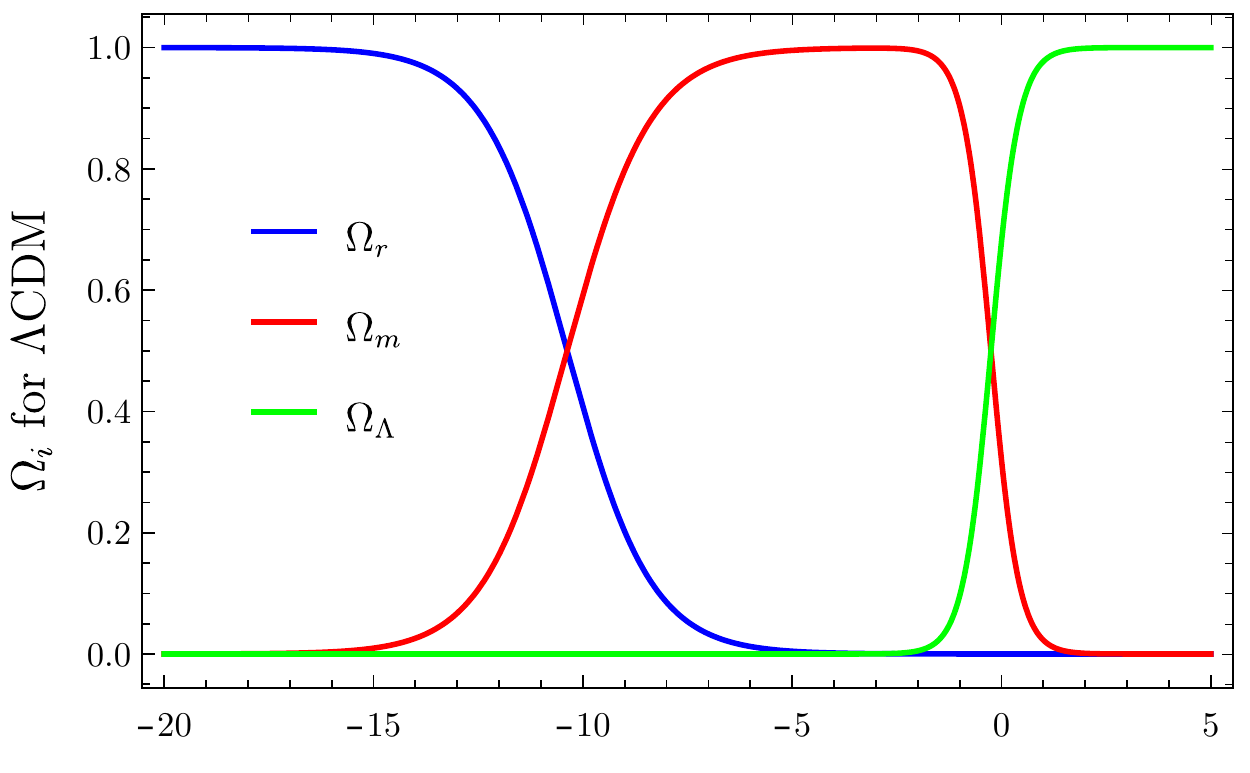}
	\includegraphics[width=.6\textwidth]{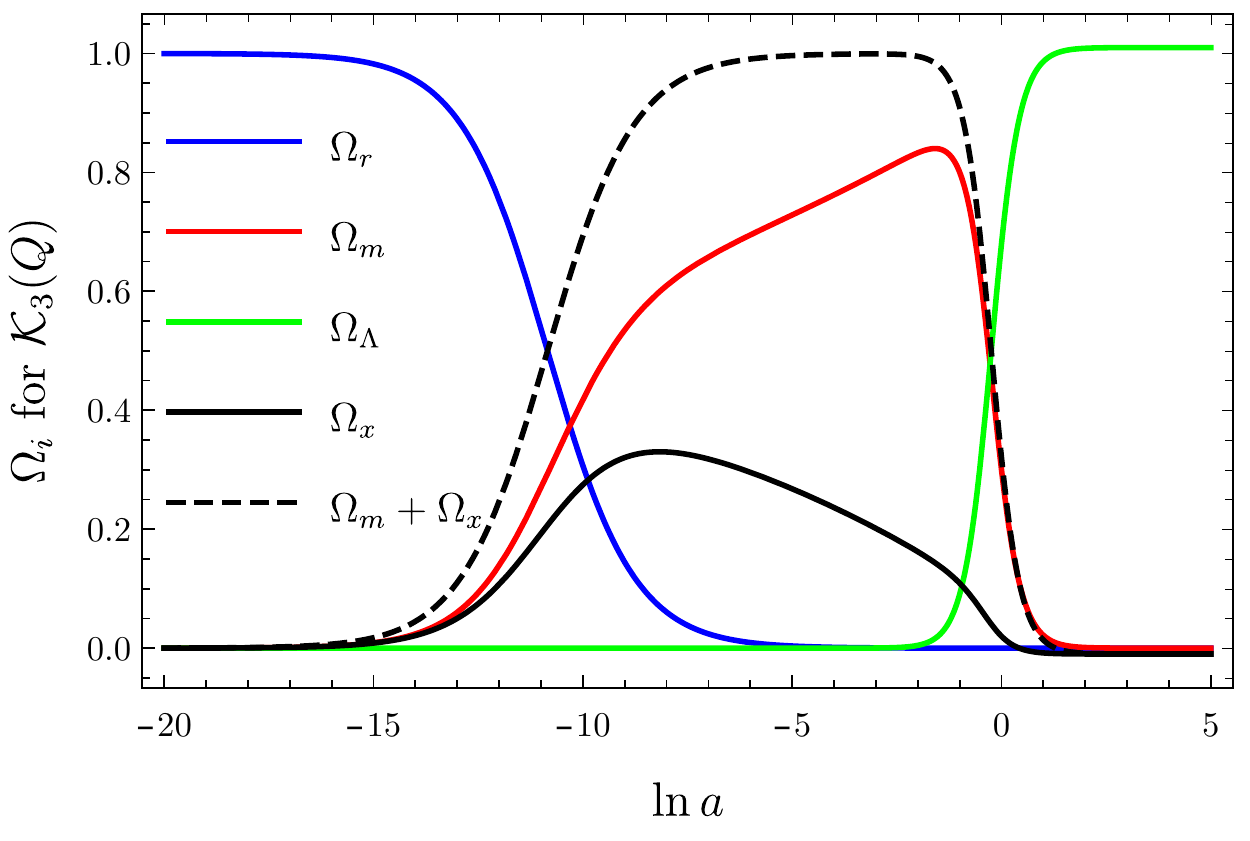}
	\caption{\label{omega_2} The evolution of the cosmic density parameters and the corresponding effective equation of state parameter in $\mathcal{K}_3$ model. The middle panel belongs to cosmic density parameters in $\Lambda$CDM. The initial conditions for the $\mathcal{K}_3$ model at $\ln a=0$ are $\Omega_m=0.3$, $\Omega_r=10^{-5}$, $\Omega_y=10^{-2}$, and $\Omega_{\Lambda}=0.68$.}
\end{figure}

In most of the figures reported so far, we have used $\Omega_y=10^{-10}$ at $\ln a=0$. It should be mentioned that choosing this very small value is not necessary. In other words, the results are not sensitive to the current value of $\Omega_y$ in the sense that it can be varied in a wide interval. As an example, in Fig.~\ref{omega_2}, we have shown the evolution of the density parameters as well as $\omega_{\text{eff}}$ in $\mathcal{K}_3$ model and $\Lambda$CDM when $\Omega_y=10^{-2}$. In the top panel, $\omega_{\text{eff}}$ is compared in $\mathcal{K}_3$ and $\Lambda$CDM models. In the middle panel, the time evolution of $\Omega$'s has been shown for the $\Lambda$CDM case. Accordingly, the bottom panel belongs to $\mathcal{K}_3$. It is clear that the deviation from the standard case is very small as far as the expansion rate is concerned. On the other hand, although the evolution of $\Omega_m$ is different in both models, the combination of $\Omega_m$ and $\Omega_x$, namely the dashed curve in the bottom panel, in the $\mathcal{K}_3$ model mimics the behavior of $\Omega_m$ in the standard model. It turns out that choosing larger values for $\Omega_y$ leads to the wrong evolution. 

\section{Discussion and conclusions}\label{CO}
In this paper, we investigated the cosmological dynamics of the newly proposed relativistic theory for MOND.
The theory has two extra fields compared to GR. A scalar field $\phi$ and a vector field $A^{\mu}$ which are supposed to play the role of the dark matter component in $\Lambda$CDM. In addition, there is a free function in the action of the theory with four different proposed forms in connection with the CMB observations \cite{Skordis:2020eui}. To study the cosmological evolution within this theory, we used the dynamical system approach. By converting the governing equations to a set of first-order differential equations, we constructed a five-dimensional dynamical system which has three  fixed points and one matter-dominated fixed line. We investigated the stability of the points and line by the linear stability theory (Table \ref{tab1}).
Using this approach, we proved that the theory has a true sequence of the cosmological epochs. The expansion rate of the cosmos in the different phases can exactly coincide with that in standard cosmology, provided that a suitable set of initial conditions is imposed. However, it is necessary to mention that the theory provides a richer structure in the sense that there are more fixed points compared to $\Lambda$CDM. Consequently, different initial conditions, in principle, could cause deviations from the standard model of cosmology. 

We reiterate that this theory has an extra fixed point ($\mathcal{P}_4$) compared to GR. This point corresponds to the $\Omega_x$ dominated phase, which does not exist in the standard cosmology. This fixed  point is unstable in all the models, and the trajectories in the phase space do not necessarily pass this nonstandard extra point except in one of the models, namely $\mathcal{K}_1$, in which $\mathcal{P}_4$ unavoidably appears at the early universe and causes violent deviation from the standard cosmology. So this specific model is certainly ruled out. This is not new and has been already noticed in \cite{Skordis:2020eui} using different interpretations. 

Our results imply that RMOND has a simple and viable cosmological behavior at the background level. More specifically, with a suitable choice of initial conditions, RMOND successfully recovers $\Lambda$CDM, while deviations from $\Lambda$CDM is possible as well.

\acknowledgments

We are very grateful to Benoit Famaey for providing us with the history of the covariant formulations of MOND. We appreciate Pavel Kroupa and Indranil Banik for their comments on the early version of this paper. Also, we thank Mordehai Milgrom, Stacy McGaugh, and Federico Lelli for valuable discussions. This work is supported by Ferdowsi University of Mashhad under Grant No. 56145 (13/09/1400).

\end{document}